\documentclass[conference]{IEEEtran}
\IEEEoverridecommandlockouts
\usepackage{amsmath,amsfonts}
\usepackage{algorithmic}
\usepackage{algorithm}
\usepackage{array}
\usepackage[caption=false,font=normalsize,labelfont=sf,textfont=sf]{subfig}
\usepackage{textcomp}
\usepackage{stfloats}
\usepackage{url}
\usepackage{verbatim}
\usepackage{graphicx}
\usepackage[top=0.7in, left=0.673in, right=0.7in, bottom=1.013in]{geometry}
\usepackage{cite}
\usepackage{hyperref}
\usepackage{pgfplots}
\usepackage{booktabs}
\pgfplotsset{compat=1.18}
\begin{document}

\title{Graph-Based Semantic Encoder–Decoder Framework for Task-Oriented Communications in Connected Autonomous Vehicles}

\author{\IEEEauthorblockN{Soheyb Ribouh}
\IEEEauthorblockA{\textit{ Univ Rouen Normandie, INSA Rouen Normandie} \\
\textit{Université Le Havre Normandie, Normandie Univ}\\
\textit{LITIS UR 4108, F-76000 Rouen, France} \\
}
\and
\IEEEauthorblockN{Phil Polo Ditsia Di Ngoma }
\IEEEauthorblockA{\textit{ Univ Rouen Normandie, INSA Rouen Normandie} \\
\textit{Université Le Havre Normandie, Normandie Univ}\\
\textit{LITIS UR 4108, F-76000 Rouen, France} \\
}

\thanks{The source code is available at: \url{https://github.com/philpolo/gbsed}}

\thanks{Accepted at IEEE International Conference on Machine Learning for Communication and Networking (ICMLCN 2026)}
}



\maketitle

\begin{abstract}
Connected autonomous vehicles (CAVs) require reliable and efficient communication frameworks to support safety-critical and task-oriented applications such as collision avoidance, cooperative perception, and traffic risk assessment. Traditional communication paradigms, which focus on transmitting raw bits, often incur excessive bandwidth consumption and fail to preserve the semantic relevance of transmitted information. To bridge this gap, we propose a Graph-Based Semantic Encoder-Decoder (GBSED) architecture tailored for task-oriented communications in CAV networks. The encoder leverages scene graphs to capture spatial and semantic relationships among road entities, combined with a semantic compression algorithm that reduces the size of the extracted graph-based representations by up to $99\%$ compared to raw images, while the decoder reconstructs task-relevant representations rather than raw data. This design enables a significant reduction in communication overhead while maintaining high semantic fidelity, exceeding $0.9$ at SNR levels above $10$~dB, for downstream vehicular tasks.  We evaluate the proposed framework through simulations in autonomous driving scenarios, where the semantic encoder and decoder are integrated into a MIMO–OFDM physical layer system. The results demonstrate high prediction success rates for risk assessment, improved robustness under the 3GPP CDL channel, and significant compression gains, confirming that the proposed semantic communication framework is a promising solution for future 6G systems.

\end{abstract}


\section{Introduction}
6G communication is expected to push the boundaries of wireless communication by exploring new paradigms that go beyond traditional methods of data transmission. This next-generation communication technology will support a wide array of innovations and  transforming
industries such as autonomous driving, augmented reality, remote surgery, holographic communication and advanced Internet of vehicles (IoV) applications\cite{dang2020should} \cite{ribouh2020multiple}. 
As connectivity in emerging technologies continues to expand, the demand for efficient and intelligent wireless networks becomes increasingly critical. Semantic communication (SC) is seen as a promising solution to address these demands by focusing on conveying the meaning behind information, rather than transmitting raw data \cite{lokumarambage2023wireless} \cite{Hu2024}. This approach offers significant advantages in terms of bandwidth efficiency and the ability to handle large volumes of content, making it particularly suitable for 6G networks and beyond. By leveraging context-aware information, SC enables intelligent systems to transmit only the most relevant data, optimizing content transmission and reducing unnecessary communication overhead. Semantic communication is expected to revolutionize the physical layer of wireless communication systems, offering high data rates with reduced bandwidth consumption \cite{ribouh2024seecad}. Recently, researchers from both academia and industry have shown increasing interest in semantic communication technologies, recognizing their potential to revolutionize wireless connectivity. By embedding intelligence not only within the network but also at the edge, this approach can significantly contribute to enabling agent-based AI \cite{Eldeeb2024}.

The concept of semantic communication was explored early on, where the authors in~\cite{guler2018semantic} proposed a semantic communication framework for text transmissions. They investigated the conditions of Bayesian Nash equilibrium by formulating the communication process as a Bayesian game, wherein actions are taken sequentially to form beliefs about the other party. The authors in~\cite{weng2021semantic} introduced DeepSC-S, a semantic communication system for speech signals. This system utilized an attention mechanism architecture based on the squeeze-and-excitation (SE) network and was designed to be adaptable across various AWGN channel conditions, making it suitable for practical multimedia transmission systems.
Recently, semantic communication has attracted significant attention, with a wide range of approaches being proposed, spanning generative models, transformer-based architectures, knowledge-graph integration, and task-oriented designs. For instance, a generative approach was adopted in~\cite{Yuan2025}, where the authors designed a multi-task generative semantic communication system capable of jointly supporting image reconstruction and image segmentation tasks from the same transmitted semantic representation. Furthermore, the authors in~\cite{Ren2025} proposed a prompt-based generative semantic communication architecture, where textual prompts can dynamically adjust what semantic content is transmitted. The work in~\cite{Li2025} extended this direction with goal-oriented wireless video transmission using generative AI, while resource allocation for energy-efficient image semantic communication was addressed in~\cite{Hu2025}.
Task-oriented designs are also emerging. Knowledge sharing in the Internet of Vehicles (IoV) via deep semantic communication was explored in~\cite{Wang2025IoV}, where the authors proposed a semantic communication framework (SCKS) targeting the sharing of neural network models (knowledge) among vehicles and RSUs in the IoV context.
In~\cite{Chen2025}, the authors introduced large language model (LLM)-based semantic communication, which maps visual features to textual or semantic descriptions, transmits those, and uses an LLM at the receiver to reconstruct or infer underwater scene imagery. Another approach based on LLMs was proposed in~\cite{ribouh2025large}, where the authors developed a semantic encoder for image-to-text conversion and employed a semantic decoder based on a stable diffusion model for image reconstruction. Beyond these, the work in~\cite{Mohsin2025} proposed a semantic communication framework based on Vision Transformers (ViTs) as both encoder and decoder. The architecture efficiently maps input images into compact semantic embeddings at the transmitter and accurately reconstructs them at the receiver under fading and noisy channels.
Finally, researchers in~\cite{Hello2024} combined semantic and pragmatic aspects by leveraging recent advances in LLMs to generate compact knowledge representations suitable for transmission between intelligent agents. They employed a cascade of LLMs and graph neural networks (GNNs) to design the semantic encoder, enabling the selection and encoding of only the most meaningful information for the receiver.

Despite these advances, most prior works have been evaluated on small datasets under simplified channel models such AWGN channel and fading channels. Few have directly addressed task-oriented vehicular scenarios, where communication efficiency, semantic fidelity, and robustness must be co-optimized to support safety-critical decision making. This gap motivates us  to  propose a  \textbf{Graph-Based Semantic Encoder–Decoder (GBSED)} framework, designed for task-oriented communication in connected  autonomous vehicle  networks. The key contributions of our work are as follows:  
\begin{itemize}
    \item  We design GBSED to leverage \emph{scene graphs} that explicitly capture semantic and spatial relationships among road entities. 

    \item We proposed a new semantic compression algorithm reduces  significantly the size of extracted graph representations compared to raw images, while preserving high semantic fidelity:

    \item \textbf We designe a Task-oriented semantic decoder regenerates task-relevant graph-based representations, thereby optimizing performance for risk prediction in CAV.

    \item  We integrate GBSED into a \emph{MIMO–OFDM}-based realistic physical-layer communication system and evaluate its robustness under \emph{3GPP CDL} channels, demonstrating high prediction success rates.
\end{itemize}

\section{System model}

\begin{figure*}[ht]
    \centering
    \includegraphics[width=1\linewidth,height=0.44\linewidth]{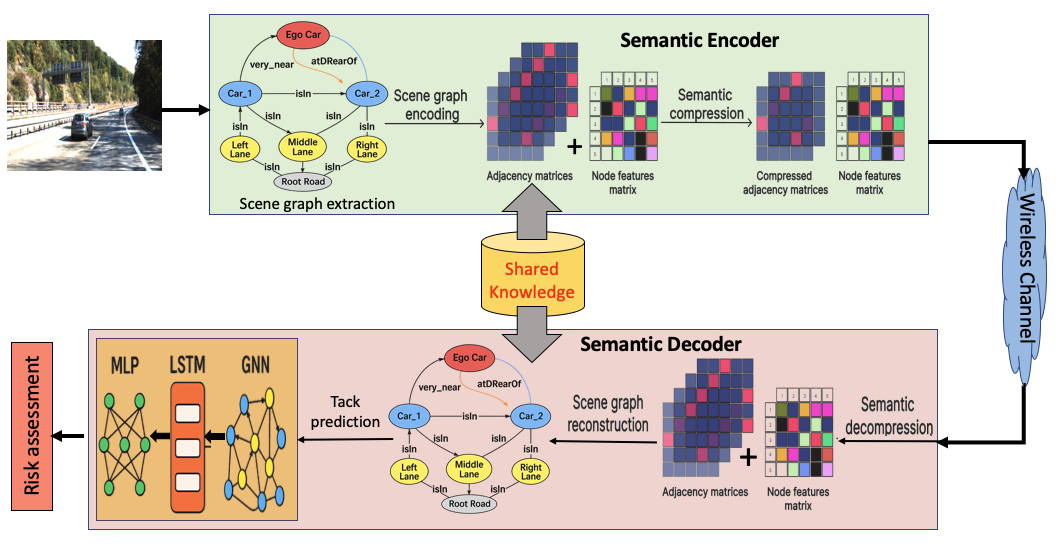}
    \caption{The proposed Graph-Based Semantic Encoder–Decoder (GBSED) architecture. }
    
    \label{fig:sg_autoencoder}
\end{figure*}

The proposed Graph-based Semantic Encoder-Decoder (GBSED) framework facilitates efficient encoding and decoding of road scene information for autonomous vehicle communication. By encoding the semantic representations of road scenes, the transmitter (e.g., an autonomous vehicle) conveys task-relevant knowledge to the receiver. This enables the receiver to perform accurate decision-making without requiring the transmission of raw data. The proposed system architecture consists of two core components:  a semantic encoder that extracts semantic representation of the road scene, and  a semantic decoder that interprets the received representation to support decision-making, as illustrated in Figure. \ref{fig:sg_autoencoder}. Both components rely on   shared knowledge, stored in a common configuration file, which specifies the relation types and node attribute indices to ensure synchronized interpretation between transmitter and receiver.

\subsection{Semantic encoder}
The semantic encoder maps an input image $\mathbf{I}$ into a structured adjacency tensor $T$ and a node feature matrix $F$, i.e.,  
\begin{equation}
    (T, F) = f_{\text{enc}}(\mathbf{I}), \quad 
    T \in \{0,1\}^{N \times N \times |\mathcal{R}|}, \quad F \in \mathbb{R}^{N \times d},
\end{equation}
where $N$ is the number of detected objects, $|\mathcal{R}|$ is the number of possible relation types, and $d$ is the feature dimension of each node.  

The semantic encoder includes the following components:  

\subsubsection{Scene graph extraction}
A scene graph provides a structured representation of visual knowledge, encoding both spatial relationships and semantic interactions among road entities. Such a representation is particularly important in autonomous driving, as it enables a compact and task-relevant abstraction of the environment.  

The extraction of a scene graph from an input image $\mathbf{I}$ can be formalized in three stages:  

\begin{enumerate}  
    \item \textbf{Object Detection:} The image is processed by a detection network $f_{\text{det}}(\cdot)$ to localize and classify road entities. Each detected object $o_i$ is characterized by a bounding box $b_i \in \mathbb{R}^4$ and a category label $c_i \in \mathcal{C}$:  
    \begin{equation}  
        \{(b_i, c_i)\}_{i=1}^N = f_{\text{det}}(\mathbf{I}),  
    \end{equation}  
    where $N$ denotes the number of detected objects.  

    \item \textbf{Inverse Perspective Mapping (IPM):} Each bounding box is projected from the image plane into a bird’s eye view (BEV) coordinate system using a geometric transformation $\mathcal{T}_{\text{IPM}}$:  
    \begin{equation}  
        p_i = \mathcal{T}_{\text{IPM}}(b_i), \quad p_i \in \mathbb{R}^2,  
    \end{equation}  
    where $p_i$ denotes the ground-plane position of object $o_i$.  

    \item \textbf{Scene Graph Construction:} A graph $\mathcal{G} = (\mathcal{V}, \mathcal{E})$ is constructed, where the node set $\mathcal{V} = \{(p_i, c_i)\}_{i=1}^N$ encodes the detected entities and the edge set $\mathcal{E}$ captures semantic relations. An edge $e_{ij} \in \mathcal{E}$ is established between nodes $i$ and $j$ according to a relational function $r(\cdot)$, which may depend on spatial proximity or predefined interaction rules:  
    \begin{equation}  
        e_{ij} = r\big((p_i, c_i), (p_j, c_j)\big).  
    \end{equation}  
\end{enumerate}  

The resulting graph $\mathcal{G}$ thus encodes both object-level attributes and inter-object semantics, providing a structured abstraction of the road scene suitable for downstream autonomous driving tasks.

\subsubsection{Scene graph encoding}
Participants in a visual scene often play diverse roles; for instance, a vehicle operates within a lane, on a road. To capture such interactions, a scene graph is modeled as a directed multi-graph $\mathcal{G} = (\mathcal{V}, \mathcal{E}, \mathcal{R})$, where $\mathcal{V}$ is the set of nodes corresponding to detected objects, $\mathcal{E}$ is the set of edges, and $\mathcal{R}$ denotes the set of relation types. Each edge $e_{ij}^r \in \mathcal{E}$ encodes a specific relationship $r \in \mathcal{R}$ between nodes $i$ and $j$, such as ``is in,'' ``is near,'' or ``to the left of''. Given the extracted scene graph, the relational structure can be represented in matrix form. 
For each relation type $r \in \mathcal{R}$, define an adjacency matrix $\mathbf{A}^{(r)} \in \mathbb{Z}^{N \times N}$ as :
\begin{equation}
    \mathbf{A}^{(r)}_{ij} =
    \begin{cases}
        r, & \text{if an edge of type } r \text{ exists from node } i \text{ to node } j,\\
        0, & \text{otherwise},
    \end{cases}
\end{equation}
where $N = |\mathcal{V}|$ is the number of nodes in the scene graph. 

To jointly encode multiple relation types, the adjacency matrices are stacked into a third-order tensor as follows:  

\begin{equation}
    T = \big[ \mathbf{A}^{(1)}, \mathbf{A}^{(2)}, \ldots, \mathbf{A}^{(|\mathcal{R}|)} \big]
    \in \{0,1,\ldots,|\mathcal{R}|\}^{N \times N \times |\mathcal{R}|},
\end{equation}
where each adjacency matrix satisfies $\mathbf{A}^{(r)}_{ij} \in \{0,r\}$.

\subsubsection{semantic compression}
As defined previously, objects within a visual scene are not uniformly linked by all possible relations, and not every potential relationship is represented in the constructed scene graph. For instance, if no car is detected to the left of the ego vehicle, the corresponding spatial relation ``to the left of'' will not appear in the graph. To address sparsity across relations, we propose a semantic compression scheme that removes adjacency matrices associated with non-existent relations in the scene graph. This selective removal yields an average compression rate of $67\%$ for the overall set of adjacency matrices, significantly reducing storage requirements and computational overhead while preserving the essential relational information present in the scene.

The semantic compression procedure is detailed in Algorithm \ref{alg:sem_comp}.The algorithm refers to the shared knowledge configuration file to ensure that the adjacency matrices in $T$ are ordered according to the same relation indices $r$ at both transmitter and receiver.  The proposed compression scheme operates with linear complexity $\mathcal{O}(|\mathcal{R}|)$ with respect to the number of defined relation types.  

\begin{algorithm}[h]
    \caption{Semantic Compression via Relation-Based Selection}
    \label{alg:sem_comp}
    \textbf{Input:} Adjacency tensor $T \in \{0,1,\ldots,|\mathcal{R}|\}^{N \times N \times |\mathcal{R}|}$ \\
    \textbf{Output:} Compressed tensor $T'$ 
    \begin{algorithmic}[1]
        \STATE \textbf{Initialization:} $T' \gets \emptyset$
        \FOR{$r = 1$ \TO $|\mathcal{R}|$}
            \STATE $\mathbf{A}^{(r)} \gets T[:,:,r]$  
            \IF{$\exists\,(i,j): \mathbf{A}^{(r)}[i,j] > 0$}
                \STATE $T' \gets T' \cup \{ \mathbf{A}^{(r)} \}$ 
            \ENDIF
        \ENDFOR
        \RETURN $T'$
    \end{algorithmic}
\end{algorithm}

\subsection{Semantic decoder}
The semantic decoder reconstructs a scene graph $\mathcal{G}$ from the compressed adjacency tensor $T'$ and the node feature matrix~$F$. 
\begin{equation}
    \mathcal{G} = f_{\text{dec}}(T', F),
\end{equation}
where $T' \subseteq T$ denotes the compressed version of the adjacency tensor.
The semantic decoder includes the following components:

\subsubsection{Semantic Decompression}  

At the receiver, the binary adjacency tensor $T$ is reconstructed as described in Algorithm~\ref{alg:sem_decomp}. First, all relation slots are reinserted: for each received adjacency matrix $\mathbf{A}^{(r)} \in T'$, the decoder places it in the $r$-th position of $T$, while all missing positions are filled with zero matrices. to ensure consistency, the decoder refers to the shared knowledge configuration file, ensuring that the adjacency matrices are ordered according to the same relation indices $r$ as those used at the transmitter. At this stage, entries remain in $\{0,r\}$. Then, the entire tensor is binarized so that each adjacency matrix becomes purely binary, with all nonzero entries mapped to $1$:

\begin{algorithm}[h]
    \caption{Semantic Decompression via Self-Describing Adjacency Matrices}
    \label{alg:sem_decomp}
    \textbf{Input:} Compressed tensor $T'$.\\
    \textbf{Output:} Binary adjacency tensor \\ $T \in \{0,1\}^{N \times N \times |\mathcal{R}|}$.
    \begin{algorithmic}[1]
        \STATE Initialize $T \gets 0_{N \times N \times |\mathcal{R}|}$ 
        \FOR{each adjacency matrix $\mathbf{A} \in T'$}
            \STATE $r \gets \operatorname{uniq}\{\, \mathbf{A}[i,j] : \mathbf{A}[i,j] > 0 \,\}$ 
            \STATE $T[:,:,r] \gets \mathbf{A}$ 
        \ENDFOR
        \FOR{$r = 1$ \TO $|\mathcal{R}|$}
            \STATE $T[:,:,r] \gets \mathbb{1}\!\left[T[:,:,r] > 0\right]$
        \ENDFOR
        \RETURN $T$
    \end{algorithmic}
\end{algorithm}

\subsubsection{Adjacency Tensor to Scene Graph}  

Scene graph restoration is achieved by reconstructing relational triplets of the form $(N_i, r, N_j)$, where $N_i$ denotes the source node, $N_j$ the target node, and $r \in \mathcal{R}$ is the relation type. These triplets are inserted into the directed multi-relational graph $G$ to recover its full structure. The reconstruction procedure is outlined in Algorithm~\ref{alg:sg_regeneration}. The computational complexity of this process is $\mathcal{O}(|\mathcal{R}| \times N^2)$, where $|\mathcal{R}|$ is the number of relation types and $N$ is the number of nodes.  

\begin{algorithm}[h]
    \caption{Scene Graph Regeneration}
    \label{alg:sg_regeneration}
    \textbf{Input:} Adjacency tensor $T \in \{0,1\}^{N \times N \times |\mathcal{R}|}$, 
    node feature matrix $F \in \mathbb{R}^{N \times d}$, 
    relation ontology $\mathcal{R}$ (pre-shared). \\
    \textbf{Output:} Scene graph $G$.
    \begin{algorithmic}[1]
        \STATE \textbf{Initialization:} $G \gets \emptyset$
        \STATE \textbf{Create nodes:} 
        \FOR{$j = 1$ \TO $N$}
            \STATE $attr_j \gets F[j]$
            \STATE $N_j \gets$ create node with attributes $attr_j$
            \STATE insert $N_j$ into $G$
        \ENDFOR
        \FOR{$r = 1$ \TO $|\mathcal{R}|$}
            \STATE $\mathbf{A}^{(r)} \gets T[:,:,r]$ 
            \STATE $rel \gets \mathcal{R}[r]$ \COMMENT{relation name from ontology}
            \FOR{$j = 1$ \TO $N$}
                \FOR{$k = 1$ \TO $N$}
                    \IF{$\mathbf{A}^{(r)}[j,k] = 1$}
                        \STATE $t \gets (N_j, rel, N_k)$ 
                        \STATE insert $t$ into $G$
                    \ENDIF
                \ENDFOR
            \ENDFOR
        \ENDFOR
        \RETURN $G$
    \end{algorithmic}
\end{algorithm}

\subsubsection{Task predilection} \label{Risk_model}
This work addresses the challenge of efficient risk assessment in autonomous connected vehicles. To this end, the reconstructed scene graph $\mathcal{G}$ is utilized as input to a predictive model. A Graph Neural Network (GNN) is first employed to extract salient structural features from $\mathcal{G}$, producing a compact vector representation. This representation is then processed by a Long Short-Term Memory (LSTM) network for temporal modeling, followed by a Multilayer Perceptron (MLP) for sequence classification, in order to predict the occurrence of potential risks.

\section{Experimental Setup}

This section details the dataset used, the training configuration used for performance assessment and the communication scenario considered for transmission.

\subsection{Dataset}
The 1043-syn dataset, introduced in \cite{Malawade}, was employed for experimental evaluation. This dataset comprises 1,044 synthetic lane-change scenarios generated using the CARLA simulator \cite{Dosovitskiy}, with a total of 50,699 images of resolution $720 \times 1280$ pixels.

\subsection{Training Configuration}
The task prediction model was trained with a batch size of 16 for 200 epochs using the AdamW optimizer with a learning rate of $5 \times 10^{-5}$. Cross-entropy loss was employed as the objective function. All experiments were conducted on an NVIDIA Tesla V100 PCIe GPU with 32 GB memory.

\subsection{Communication scenario} The semantic encoder and decoder are integrated into the  MIMO–OFDM physical layer system introduced in \cite{NRX}. The system configuration is as follows: the resource grid consists of $132$ subcarrier slots, each spaced at $240$ kHz, with $14$ OFDM symbols per frame and data symbols are mapped using 64-QAM modulation. We consider a $2 \times 4$ MIMO configuration, where the resource grid is transmitted through two transmit antennas and received at four antennas via the 3GPP CDL channel model.

\section{Results and Discussion}

The performance of the proposed framework was evaluated using complementary metrics. Communication efficiency was quantified through the compression rate. System reliability was assessed by measuring semantic fidelity, which captures the ability to preserve the original meaning of the transmitted data after reconstruction. Finally, the overall effectiveness of the system was validated on the downstream risk classification task, using accuracy, precision, and recall as standard indicators.
\subsection{Data Compression }
The Compression Ratio (CR) is defined as the ratio between the size of the raw scene images and their encoded representation at the output of the semantic encoder.  

As shown in Table \ref{tab:compression}, the proposed GBSED method achieves a remarkable compression gain. Specifically, the exchanged data volume is reduced from $13$~GB of raw image data to only $5.37$~MB, which corresponds to a data reduction of approximately $99.9\%$ (i.e., a compression ratio of $2425$). In addition, relative to JPEG compression, which reduces the raw data volume by $59.5\%$, the proposed method still achieves a highest data reduction, underlining the significant advantage of semantic communication for efficient data transmission.

\begin{table}[h]
\centering
\caption{Comparison size, compression ratio, and reduction across methods.}
\begin{tabular}{lccc}
\toprule
\textbf{Method} & \textbf{Size / 5060 images} & \textbf{CR } & \textbf{Reduction (\%)} \\
\midrule
Raw RGB (24 bits)   & 13.0 GB  &  1     & 0 \% \\
JPEG (measured)       & 5.29 GB & 2.47    & 59.5 \% \\
\textbf{Proposed GBSED }  & \textbf{5.37 MB }  & \textbf{2425}    & \textbf{99.9 \%} \\
\bottomrule
\end{tabular}
\label{tab:compression}
\end{table}

\subsection{Performance Analysis of Task-oriented Semantic Fidelity}

The proposed framework was evaluated under a 3GPP CDL channel model across different signal-to-noise ratio (SNR) values ranging from $0$~dB to $20$~dB. To quantify system reliability, we measured semantic fidelity, defined as the degree to which the meaning intended by the transmitter is preserved at the receiver. It is computed as the ratio of correctly recovered semantic entities (objects and relations in the scene graph) to the total number of transmitted semantic entities. This metric captures the extent to which the reconstructed graph supports reliable downstream task execution.

\begin{figure}[ht!]
    \centering
    \includegraphics[width=1.0\linewidth,height=0.19\textheight]{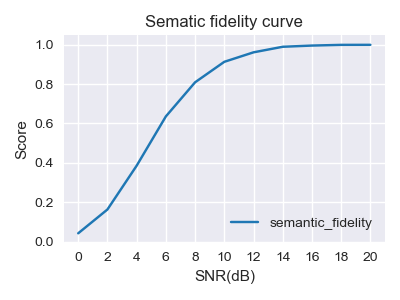}
    \caption{Semantic fidelity as a function of the Signal-to-Noise Ratio (SNR).}
    \label{fig:sem_fidelity}
\end{figure}

Fig.~\ref{fig:sem_fidelity} illustrates the semantic fidelity performance of the proposed graph-based semantic communication framework under varying SNR values. As observed, semantic fidelity remains very low for SNR values below $4$~dB, indicating that the recovered scene graphs are highly corrupted and fail to capture the essential objects and relations required for reliable task execution. Between $4$~dB and $8$~dB, semantic fidelity improves sharply, reaching values above $0.8$ at an SNR of $8$~dB, which corresponds to the preservation of most task-relevant semantic structures. For SNR values greater than $10$~dB, semantic fidelity exceeds $0.9$, demonstrating that the reconstructed graphs closely match the transmitted semantics. This high-fidelity regime ensures reliable task-oriented communication, as nearly all critical objects and relations are successfully preserved at the receiver. Beyond $14$~dB, semantic fidelity saturates near $1.0$, indicating almost perfect semantic recovery.
These results confirm that the proposed framework achieves robust task-oriented communication when operating at moderate-to-high SNR levels, with reliable downstream performance guaranteed once semantic fidelity surpasses the $0.9$ threshold.

\subsection{Risk Assessment Performance Evaluation}
\begin{figure}[t]
    \centering
    \includegraphics[width=1.0\linewidth,height=0.25\textheight]{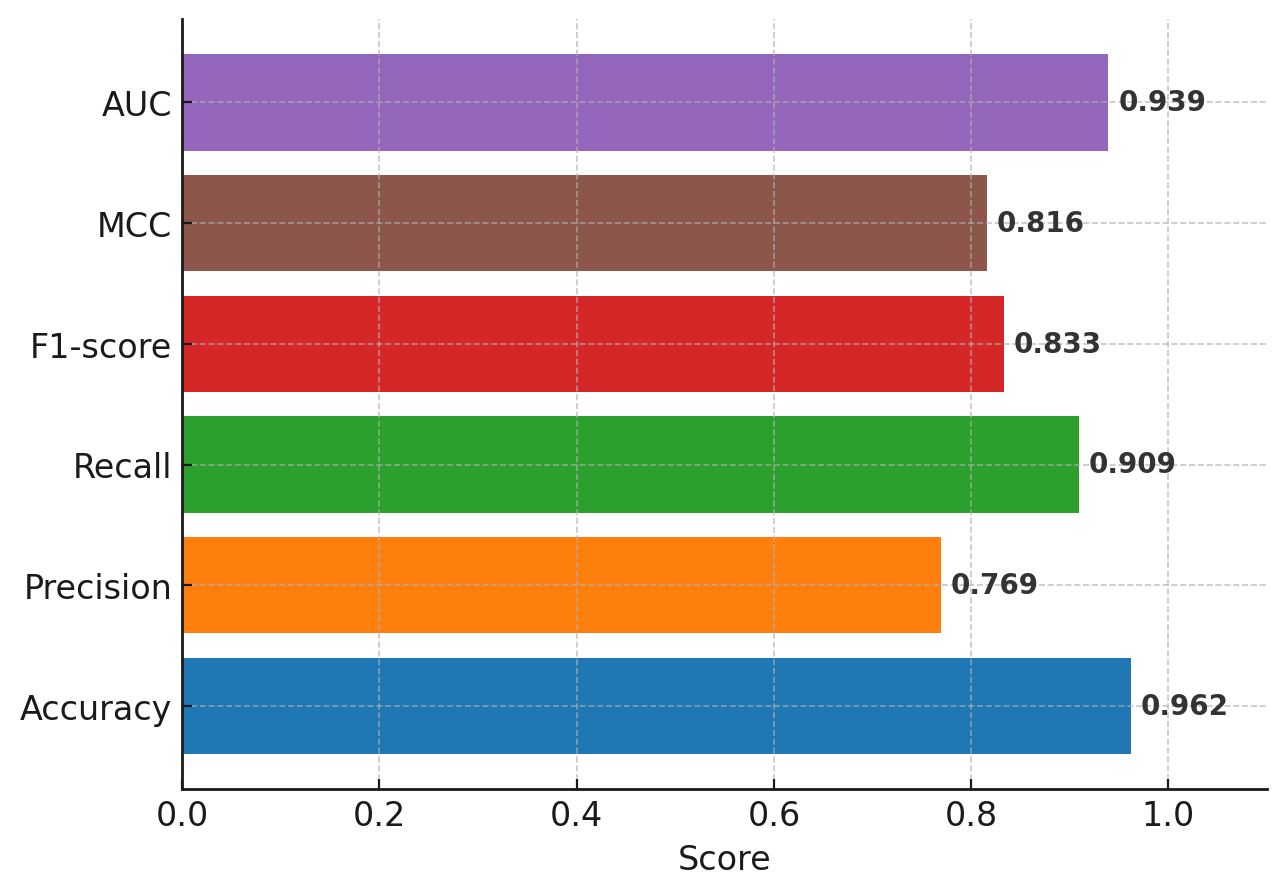}

\caption{Risk Assessment Model Performance (1043-syn Dataset \cite{Yu})}
\label{fig:risk-perf-1043}
\end{figure}

To validate the capability of the proposed system in supporting task-oriented communication, We compute the downstream performance of the proposed framework for risk assessment across multiple evaluation metrics, including accuracy, precision, recall, F1-score, AUC, and the Matthews correlation coefficient (MCC) \cite{powers2011evaluation}. \\
 The results indicate that the risk prediction model (detailed in subsection \ref{Risk_model}) achieves high overall reliability, with an accuracy of $0.962$, AS shown in Fig.~\ref{fig:risk-perf-1043}. The achieved Precision is $0.769$, reflecting the presence of some false positives, whereas recall achieves a significantly higher value of $0.909$, confirming the model’s ability to successfully capture most true risk events. The F1-score is $0.833$ demonstrates a balanced trade-off between precision and recall, ensuring both sensitivity and reliability in risk detection. The AUC score of $0.939$ highlights the excellent discriminative capability of the system in distinguishing risk from non-risk scenarios. Finally, the MCC reaches $0.816$, validating that the predictions are well-balanced across both classes.
These results confirm that the proposed semantic communication framework ensures robust and reliable risk assessment.\\

\begin{figure}
    \centering
    \includegraphics[width=1\linewidth,height=0.3\textheight]{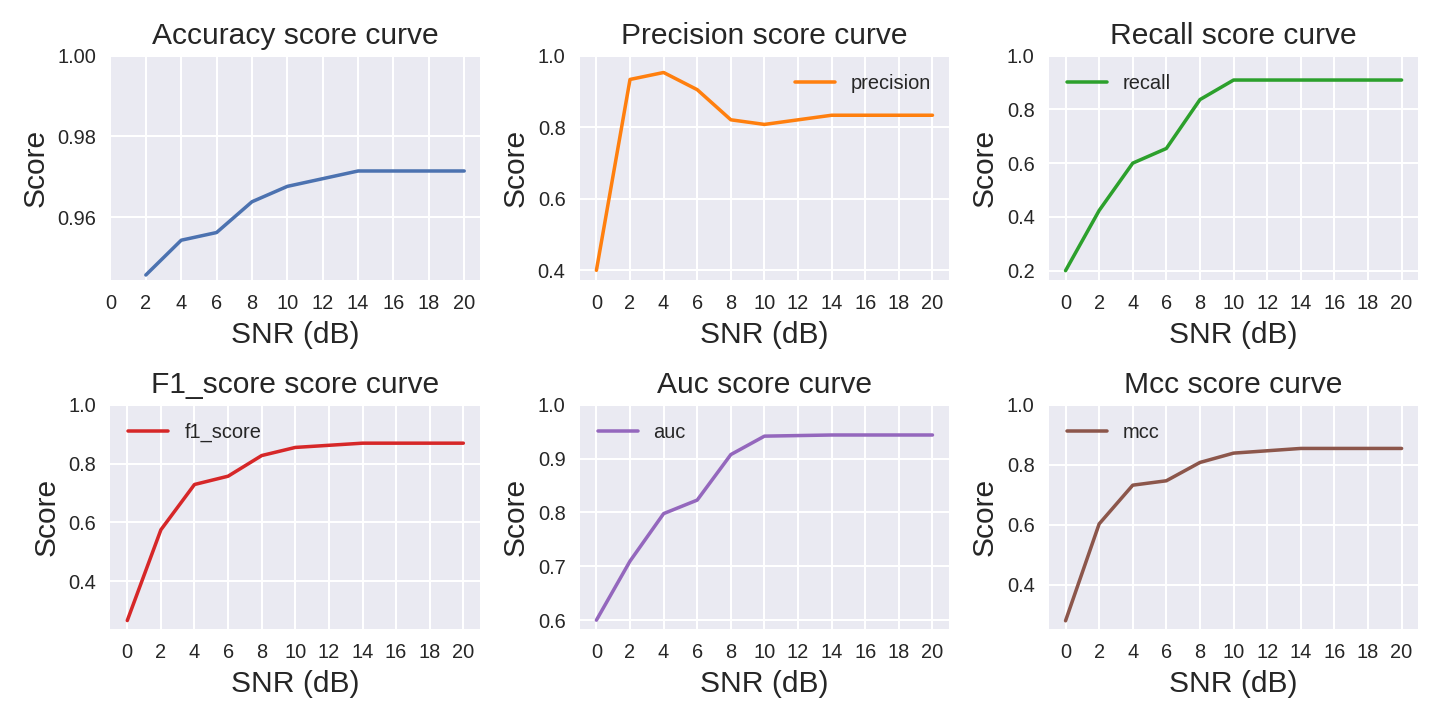}
    \caption{ Performance metrics of the proposed model as a function of the Signal-to-Noise Ratio (SNR). The plots show the model's accuracy, precision, recall, F1-score, Area Under the Curve (AUC), and Matthews Correlation Coefficient (MCC).}
    \label{fig:performance}
\end{figure}

Fig.~\ref{fig:performance} presents the task performance evaluation metrics across different SNR values. The results show that task accuracy remains consistently above $0.95$ for all SNR levels, highlighting the robustness of the proposed framework to channel degradation. However, accuracy alone masks critical variations in other metrics.
At low SNR ($0$–$4$~dB), recall, F1-score, AUC, and MCC remain below $0.6$, indicating that many relevant semantic entities are missed despite high overall accuracy. As SNR increases beyond $8$~dB, these metrics improve sharply, with recall and F1-score exceeding $0.8$, where AUC and MCC approaching $0.9$. This regime corresponds to semantic fidelity values above $0.8$, where most task-relevant objects and relations are successfully preserved.
For SNR values $\geq 10$~dB, all metrics saturate, with recall exceeding $0.9$, AUC stabilizing near $0.95$, and MCC around $0.85$. This high-SNR regime corresponds to semantic fidelity exceeding $0.9$, ensuring that downstream risk assessment task can be executed with high reliability.

\section{Conclusion and Future Works}
This paper has presented a novel Graph-Based Semantic Encoder–Decoder (GBSED) framework for efficient information exchange in autonomous driving. The framework integrates image-based scene understanding, task-oriented risk prediction, and bandwidth-optimized semantic compression. Simulation results demonstrate that the proposed semantic communication system achieves significant bandwidth savings, with compression ratios of $2425$ compared to raw data , while maintaining high task performance and robustness under a 3GPP CDL channel with MIMO–OFDM transmission highlighting its potential as an enabling technology for future 6G systems. Future work will focus on extending the framework to support image reconstruction at the receiver, thereby further enhancing situational awareness in connected and autonomous vehicles.


\vfill

\end{document}